\documentclass{article}
\usepackage[utf8]{inputenc}
\usepackage{hyperref}
\usepackage[numbib]{tocbibind}
\usepackage{authblk}
\usepackage[
backend=biber,
style=alphabetic,
sorting=ynt
]{biblatex}
\usepackage{blindtext}
\usepackage[style=alphabetic]{biblatex}
\usepackage{hhline}
\usepackage{mathptmx}
\usepackage{subcaption}
\usepackage{amsmath}
\title{A Bayesian Learning , Greedy agglomerative clustering approach and evaluation techniques for Author Name Disambiguation Problem}
\author{Shashwat Sourav$^1$}
\date{%
    $^1$Indian Institute of Science Education and Research,Bhopal,Madhya Pradesh,India\\%
    }
\begin{document}

%\address{Department of Data Science and Engineering,Indian Institute of Sciene Education and Research (IISER),Bhopal, Madhyya Pradesh, India}

\maketitle

\section{Abstract}
Author names often suffer from ambiguity owing to the same author appearing under different names and multiple authors possessing similar names. It creates difficulty in associating a scholarly work with the person who wrote it, thereby introducing inaccuracy in credit attribution, bibliometric analysis, search-by-author in a digital library, and expert discovery. A plethora of techniques for disambiguation of author names have been proposed in the literature. I try to focus on the research efforts targeted to disambiguate author names. I first go through the conventional methods, then I discuss evaluation techniques and the clustering model which finally leads to the Bayesian learning and Greedy agglomerative approach. I believe this concentrated review will be useful for the research community because it discusses techniques applied to a very large real database that is actively used worldwide. The Bayesian and the greedy agglomerative approach used will help to tackle AND problems in a better way. Finally, I try to outline a few directions for future work.

\section{Introduction}
For any work of literature, a fundamental issue is to identify the individual(s) who wrote it, and conversely, to identify all of the works that belong to a given individual. Attribution would seem to be a simple process (putting aside those works that are published anonymously) and yet it represents a major, unsolved problem for information science. Consequently, it is necessary to analyze the metadata, and sometimes the text, of a work of literature to make an educated guess as to the identity of its authors.   Author name disambiguation comprises four distinct challenges:  First, a single individual may publish under multiple names—this includes
\begin{itemize}
    \item orthographic and spelling variants
    \item spelling errors
    \item name changes over time as may occur with marriage, religious conversion or gender re-assignment
    \item the use of pen names.  Second, many different individuals have the same name – in fact, common names may comprise several thousand individuals.
\end{itemize}
Disambiguation is needed in order to create link-outs from databases or digital libraries to online resources, including full-text papers and the authors’ home pages as well (if present). This entails knowing the individuals, not merely the names, on the paper. Thus, the rise of large bibliographic databases has invited data-mining analyses to 4
understand large-scale features of the data as a whole, and extract, re-assemble and synthesize the raw information to create entirely new knowledge. Author name disambiguation is a fundamental step in mapping knowledge domains (\cite{shiffrin2004mapping}) and in other bibliometric and scientometric analyses. It will also be useful to marketers who wish to direct their advertisements to specific individuals. Finally, I will discuss in some detail, the machine learning approach and what are the transitivity constraints while facing AND problem.
\section{Problem Statement}Author name disambiguation (AND) is one of the most vital problems in scientometrics, which has become a great challenge with the rapid growth of academic digital libraries. Existing approaches for this task substantially rely on complex clustering-like architectures, and they usually assume the number of clusters is known beforehand or predict the number by applying another model, which involves increasingly complex and time-consuming architectures. In the following section, I try to discuss some conventional methods for tackling the AMD issue and its shortcomings. I then review some of the general evaluation tecniques that are used in Machine Learning. To tackle this problem I use the Bayesian Learning and Greedy Agglomerative approach which is discussed in \ref{Section 5}.

\subsection{Why not just establish a directory of unique author identifiers?}
Before surveying current research approaches, one might ask: Why not simply set up a directory of author names with unique IDs?  A directory would, in principle, solve the problem of author name disambiguation prospectively, and if each author submitted a list of their pre-existing publications when they join the system, it would allow one to assign many articles and books retrospectively as well.  Technically, it is no more difficult to implement such a registry than to maintain any other web-based service that relies upon author registration; there are lots of public\cite{dervosinformation} which give a good overview of the security, authentication, and programming issues that are involved in this endeavor. Authors would then be responsible for using this number in all of their publications (for the rest of their lives), and it is assumed that authors will agree to remember their passwords and will update their metadata periodically (this is not mandatory, but otherwise the reliability and value of the metadata will degrade rapidly). 
\par But here we have an issue. Although the scheme has conceptual simplicity and is technically feasible, it fails to take into account the \textbf{realities of human behavior}.  Authors are not only expected to cooperate voluntarily and actively but to enter their data accurately and periodically over a ~50-year time span.  For this to work, the vast majority of authors need to participate -- even those who were only seventh-listed authors on a single article written while they were student technicians on a project.

\subsubsection{Manual Disambiguation : Pros and Cons}
Most cases to date in which author names have been disambiguated have tended to involve manual curation.  For example, librarians have traditionally carried out authority control on book collections \cite{maxwell2002maxwell}. Several initiatives make use of a combination of automatic and author-supplied or community-supplied input:  For example, there are instances in which one extracts author information from within a defined database research community, and displays it in a standardized format that is subject to manual correction. The FOAF (Friend of a Friend) initiative is a community-driven effort to define an RDF vocabulary for expressing metadata about people, and their interests, relationships, and activities. In any way, manual disambiguation is a surprisingly hard and uncertain process, even on a small scale, and is entirely infeasible for common names.  For example, in a recent study, we chose 100 names of MEDLINE authors at random, and then a pair of articles were randomly chosen for each name; these pairs were disambiguated manually, using additional information.

\section{Some Research Approaches to solve the issue of AND}
Author name disambiguation involves some of the same issues as other kinds of entity recognition and resolution. As an illustration, many active research efforts are devoted to recognizing named entities within texts and on webpages (\cite{mann2003unsupervised} \cite{cohen2005survey}), disambiguating word sense \cite{schuemie2007evaluation}, and identifying co-reference mentions. Record linkage is also a related problem
 as it involves deciding whether two different entries (in the same or different databases) refer to the same person. Nevertheless, author name disambiguation is potentially a much richer enterprise than these other tasks because it goes beyond particular mentions or particular articles to provide an overall characterization of an individual. On the other hand, named entity recognition may attempt to identify WHICH Matthew Bush is being mentioned in a particular article, author disambiguation incorporates information across all of an individual’s works, and includes features as well that involve extensive computation and outside knowledge from external sources. At the very basic level, most research approaches to author name disambiguation share the broad outlines of predictive\textbf{ “machine learning” } \cite{mitchell1997does}, which is designed to cluster or classify a body of works of literature corresponding to the individuals who wrote them. Machine learning generally requires acquiring training sets that provide
positive and negative examples; one or more features that are extracted from the works or their metadata; some decision procedure of optimization or “learning” that acts upon the features; and some means of evaluation of system performance. However, different existing systems vary to extremes in the manner in which these steps are formulated and carried out.

\subsection{The Machine Learning Approach}

Machine learning generally requires acquiring training sets that provide
positive and negative examples; one or more features that are extracted from the works or their metadata; some decision procedure of optimization or “learning” that acts upon the features; and some means of evaluation of system performance. However, different existing systems vary to extremes in the manner in which these steps are formulated and carried out. At least 10 different approaches have been described in the past few years, which will be reviewed, compared, and contrasted in this and the following section. At the outset, it is important to keep in mind that different systems should not be compared based on performance parameters (e.g., recall and precision) alone since each system was developed for a different type of disambiguation task and dataset, though generally each of the methods could potentially be applied to any of the major bibliographic databases such as DBLP, CiteSeer, arXiv, and the NASA ADS. Most disambiguation approaches fall into one of the two machine learning paradigms: supervised or unsupervised. Supervised approaches take as input a set of training examples consisting of pairs of articles that are labeled as either positive (author match)
or negative (not author match), while unsupervised approaches do not use labeled training examples. In general, supervised approaches perform better because they are tuned specifically to the data (e.g., to determine the relative importance and interactive 11 effects of different features such as a co-author vs. journal name vs. title word vs.
affiliation). Having a sufficient amount of training data is critical to the performance of any predictive model that will be extrapolated to new heretofore-unseen examples. The amount of data sufficient for training depends on the complexity of the model. Generating training sets does not have to be a manual, tedious and error-prone process; in fact, it can be done automatically too. Training sets should represent the entire database and not exhibit bias towards certain values of the predictive features (e.g., using personal email addresses will bias the dataset towards newer papers, and using suffixes will give a bias towards English names). Thus, the bias needs to be measured, and accounted for, if significant correlations with predictive features are detected. Training sets can also be generated using a hybrid of manual and automatic methods as in the paradigm of active learning (\cite{kanani2007improving},\cite{torvik2006discovering}), a strategy in
which the learning algorithm iteratively detects the most informative examples for manual curation, and the disambiguation model is updated after each iteration.  
\subsection{Evaluation Measures}
Evaluation measures from information retrieval are borrowed or adapted to evaluate the quality of the generated author individual clusters ${C_i}$. We summarise the common measures that have been defined in the literature to evaluate AND algorithms.
Let:-
\begin{itemize}
    \item TP = total number of pairs (of author names) correctly put into same cluster
    \item TN = total number of pairs correctly put into different clusters
    \item FP = total number of pairs incorrectly put into same cluster
    \item FN = total number of pairs incorrectly put into different clusters.
\end{itemize}

\begin{equation}\label{equation 1}
    \centering
    S = TP + TN + FP + FN.
\end{equation}

Here:-
\begin{itemize}
    \item Accuracy = $TP+TN/S$. It is the ratio of the number of correctly clustered pairs to the total number of pairs.
    \item  Pairwise precision pp = $TP/(TP + FP)$. It is the fraction of pairs in a cluster being co-referent.
    \item  Pairwise recall pr = $TP/(TP + FN)$. It is the fraction of co-referent pairs that are put in the same cluster.
    \item Pairwise F1-score $PF_1$ = $(2 \times pp \times pr)/(pp + pr)$. It is the harmonic mean of pp and pr.
\end{itemize}

\subsection{Designing a Clustering Model and computing it's purity}
Let's assume Suppose $M_gold$ is the number of manually generated (true or golden) author-individual clusters, $M_cor$ is the number of completely correct clusters generated by the AND algorithm and $M_gen$ is the total number of clusters generated by the AND algorithm.
Let N be the number of author references in the dataset. Let $n_{ij}$ be the total number of references in the automatically generated cluster I belonging to the corresponding manually generated cluster j. Let $n_i$ be the total number of references in the automatically generated cluster i. 
We can find the Average Cluster Purity (ACP) as :-
\begin{equation}\label{equation 2}
    \centering
     ACP = \frac{1}{N}\sum_{j=1}^{j=M_{gen}} \sum_{i=1}^{M_{gold}} \frac{n_{ij}^2}{n_{ij}} \
\end{equation}

In a similar manner we can also define the Average Author Purity (AAP) as :-
\begin{equation}\label{equation 3}
    \centering
    AAP = \frac{1}{N}\sum_{j=1}^{j=M_{gen}} \sum_{i=1}^{M_{gold}} \frac{n_{ij}^2}{n_{ij}} \
\end{equation}

From \ref{equation 2} and \ref{equation 3} we can calculate the K-Measure which is given by :-
\begin{equation}\label{equation 4}
    \centering
    K = \sqrt{ACP \times AAP}
\end{equation}
Let the given set of author references S = {${s_1 , s_2,..s_n}$} be partitioned  into a set of clusters
V = {${V_1,V_2,...V_{v}}$}.. Let the true (manually generated) set of disjoint author individual clusters be C = ${C_1,C_2,...C_c}$. Let us assume that $V_{s_{i}}$ is the automatically generated cluster to which $s_i$ belongs and $C_{s_{i}}$ denote the manually generated cluster.
Through these techniques one can evaluate the efficiency of their clustering model.
Now we can define the precision and recall of a reference $s_i$ as:-
\begin{equation}\label{equation 5}
    \centering
    p(s_i) = \frac{ |s \epsilon V(s_i) : C(s) = C_{s_{i}}|}{V(S_i)}
    %p(s_i) = \frac{ |s \epsilon C(s_i) : V(s) = V_{s_{i}}|}{C(S_i)}
\end{equation}

\begin{equation}\label{equation 6}
    \centering
    p(s_i) = \frac{ |s \epsilon C(s_i) : V(s) = V_{s_{i}}|}{C(S_i)}
\end{equation}

\section{Bayesian Learning and Greedy Agglomerative Clustering.}\label{Section 5}
Here I have tried to follow a simple hypothesis which is also followed in \cite{10.1145/1552303.1552304} that papers written by the same author should show much higher similarities in personal information of the author and other citation attributes compared to papers by different authors. First, it creates the author blocks on LN-FI author names. Given a pair of citations $p1,p2$ corresponding to two author name instances $s1,s2$ respectively in a block, it constructs a multidimensional similarity profile $x(p1, p2)$, based on title, journal name, coauthor names, MeSH, language, affiliation, email and name attributes (like the popularity of the last name, middle initial and suffix). Such profiles are computed for two reference sets: a match set M containing article pairs both of which are most likely to be written by the same individual, and a non-match set N containing article pairs known to be written by different persons. These sets are auto-generated and hence, not entirely error-free. As an outcome of the training, for each x, a value $r(x) = P(x|M)=P(x|N) (where P(x|M)$ denotes the probability of observing the similarity profile x given that the papers $p1, p2$ are by the same author and P(x|N) denotes the probability of observing x given that $p1, p2$ are by different authors) is computed, smoothed and stored. For a pair of test citations,
the similarity profile $x_test$ is computed and $r(x_test)$ is looked up (possibly extrapolated). Using a prior for match probability $P(M)$ (which varies with author names) and Bayes’s theorem, the r-value is converted to $P(M|x_test)$, the probability of match given $x_test$. This is also denoted by $P_{ij}$, the pairwise probability of citations $p_i$ and $p_j$ to belong to the same individual. However, there could be transitivity violations, that is, $P_{i, j}$, $P_{j, k}$ are high but $P_{i, k}$ is low which is unexpected.
Mathematically, this is expressed as:-
\begin{equation}\label{equation 7}
    \centering
    P_{ij} + P_{jk} -1 > P_{ik} + \Delta ; P_{ik} \geq P_{ij}; P_{jk}
\end{equation}

Here $\Delta$ is a small positive quantity. Transitivity violations are corrected by minimising the aggregate of weighted least squares defined as :-
\begin{equation}\label{equation 8}
    \centering
    W_{ij}(P_{ij} - Q_{ij})^2 + Wjk (P_{jk} - Q_{jk})^2 + W_{ik} (P_{ik }- Q_{ik} )^2
\end{equation}
Here, $P_{ij}$ is updated to $Q_{ij}$ and so
on. The weights are first set as $W_{ij} = \frac{1}{P_{ij}(1 - P_{ij})}$ and so on. Then the weights of the lower probabilities are reduced. This is similar to K-Means Clustering Algorithm that we use in Unsupervised Learning Algorithm Problems. Greedy agglomerative clustering is used;
as it does not require the number of clusters to be known beforehand. All citations are first put into singleton clusters; at each step, a pair of clusters with the largest match odds are merged. Clustering is stopped when the largest pairwise probability $P_{ij} < 0:5$. This creates high-recall clusters which contain a few mixed citations.

\section{Summary and Conclusion}\label{Section 6}
The Author's name disambiguation within bibliographic databases is a very active area of research within the computer science community. Many different features have been employed for modeling, and several quite imaginative and powerful approaches have been proposed that include higher-order comparisons among documents, groups of co-authors and other social network data, and external information obtained from web pages. The limiting performance factor is not access to enough information, but rather the computational load involved in taking all of the available information into account, which currently limits their extension to very large databases or digital libraries. 

\subsection{Challenges for Future Research}
There is no single paradigmatic author name disambiguation task – each
bibliographic database, each digital library, and each collection of publications have their own unique set of problems and issues. The collections differ in size, author diversity, and curation reliability, as well as in the types of metadata that are assigned to each publication, the cultural context of how the data are used, and the rate of growth of new items. For certain purposes (e.g., awarding the Nobel Prize to the author of a breakthrough), it may be very important to achieve high accuracy of disambiguation. For other purposes (e.g., as an aid to routine information retrieval), it may suffice to assign a high proportion of a person’s articles correctly, with little penalty occurring if
some articles are missed or misassigned. Certainly, the machine learning approach and the clustering model approach discussed need room for further improvement in precision and recall, either by encompassing additional features or by combining aspects of different models into one. However, optimizing performance is only one of the frontiers for future research. A quick-and-dirty algorithm may still be preferred over a high performing one if it is scalable, efficient, rapid, and easy to pre-compute (so that disambiguation does not need to be computed in real-time).
Each of these represents major computing challenges.

\section{Acknowledgements}\label{Section 7}
I would like to thank Dr.Parthiban Srinivasan for teaching us about the AND problem and Dr. Tanmay Basu whose Machine Learning lectures helped me in implementing the evaluation and bayesian approach to this problem. I am also grateful to all the Teaching Assistants who helped in clarifying my doubts to carry out this project smoothly.

\printbibliography

\end{document}